\documentclass[showpacs,english,aps,prc,twocolumn,superscriptaddress,floatfix]{revtex4}
\usepackage[T1]{fontenc}
\usepackage[latin1]{inputenc}
\usepackage{graphicx}
\usepackage{amssymb}
\usepackage{verbatim}
\usepackage{epic,eepic}
\usepackage{babel}

\newcommand{\ba}{\begin{eqnarray}}
\newcommand{\ea}{\end{eqnarray}}

\begin{document}

\title{New supersymmetric quartet of nuclei in the $A \sim 190$ mass region}

\author{J. Barea}
\affiliation{Center for Theoretical Physics, Sloane Physics Laboratory, 
Yale University, P.O. Box 208210, New Haven, Connecticut 06520-8120, U.S.A.
\footnote{Present address: Facultad de F{\'{\i}}sica, Universidad de Sevilla, 
Avda. Reina Mercedes s/n, E-4012 Sevilla, Spain}}
\author{R. Bijker}
\affiliation{Instituto de Ciencias Nucleares, 
Universidad Nacional Aut\'onoma de M\'exico, 
A.P. 70-543, 04510 M\'exico, D.F., M\'exico}
\author{A. Frank}
\affiliation{Instituto de Ciencias Nucleares, 
Universidad Nacional Aut\'onoma de M\'exico, 
A.P. 70-543, 04510 M\'exico, D.F., M\'exico}
\author{G. Graw}
\affiliation{Sektion Physik, Ludwig-Maximilians-Universit\" at M\" unchen, D-85748 Garching, Germany}
\author{R. Hertenberger}
\affiliation{Sektion Physik, Ludwig-Maximilians-Universit\" at M\" unchen, D-85748 Garching, Germany}
\author{H.-F.Wirth}
\affiliation{Sektion Physik, Ludwig-Maximilians-Universit\" at M\" unchen, D-85748 Garching, Germany}
\affiliation{Physik-Department, Technische Universit\" at M\" unchen, D-85748 Garching, Germany}
\author{S. Christen}
\affiliation{Institut f\" ur Kernphysik, Universit\" at zu K\" oln, D-50937 K\" oln, Germany}
\author{J. Jolie}
\affiliation{Institut f\" ur Kernphysik, Universit\" at zu K\" oln, D-50937 K\" oln, Germany}
\author{D. Tonev}
\affiliation{Institut f\" ur Kernphysik, Universit\" at zu K\" oln, D-50937 K\" oln, Germany}
\author{M. Balodis}
\affiliation{Institute of Solid State Physics, University of Latvia,
Riga, LV-1063, Latvia}
\author{J. B\={e}rzi\c{n}\v{s}}
\affiliation{Institute of Solid State Physics, University of Latvia,
Riga, LV-1063, Latvia}
\author{N. Kr\={a}mere}
\affiliation{Institute of Solid State Physics, University of Latvia,
Riga, LV-1063, Latvia}
\author{T. von Egidy}
\affiliation{Physik-Department, Technische Universit\" at M\" unchen, D-85748 Garching, Germany}

\date{\today}

\begin{abstract}
We present evidence for a new supersymmetric quartet in the $A \sim 190$ 
region of the nuclear mass table. New experimental information on transfer and neutron 
capture reactions to the odd-odd nucleus $^{194}$Ir strongly suggests the existence 
of a new supersymmetric quartet, consisting of the $^{192,193}$Os and $^{193,194}$Ir nuclei. 
We make explicit predictions for the odd-neutron nucleus $^{193}$Os, and suggest that its 
spectroscopic properties be measured in dedicated experiments. 
\end{abstract}


\maketitle

\section{Introduction}

Nuclear supersymmetry is a composite particle phenomenon that should not be confused with 
fundamental supersymmetry, as used in particle physics and quantum field theory, where it is 
postulated as a generalization of the Lorentz-Poincar\'e invariance as a fundamental symmetry 
of Nature and predicts the existence of supersymmetric particles, such as the photino and 
the selectron, for which, however, experimental evidence is yet to be found. If experiments 
about to start at the LHC at CERN find evidence of supersymmetric particles, the supersymmetry 
would be badly broken, as their masses must be much higher than those of their normal partners. 
In contrast to particle physics, nuclear supersymmetry has been subjected to experimental 
verification. 

Nuclear supersymmetry was proposed more than twenty five years ago \cite{FI} in the 
context of the interacting boson model (IBM) and the interacting boson-fermion model (IBFM) 
which have proved remarkably successful in providing a unified framework of 
even-even \cite{ibm} and odd-even nuclei \cite{ibfm}, respectively. One of its most 
attractive features is that it gives rise to a simple algebraic description, in which 
dynamical symmetries and supersymmetries play a central role, both as a way to improve 
our basic understanding of the importance of (super)symmetry in nuclear dynamics, and 
as a starting point for more precise calculations. Nuclear supersymmetry 
provides a theoretical framework in which different nuclei are treated as members of 
the same supermultiplet and whose spectroscopic properties are described by 
a single Hamiltonian and a single set of transition and transfer operators. 

Originally nuclear supersymmetry was formulated as a symmetry among pairs of nuclei 
consisting of an even-even and an odd-even nucleus \cite{FI,u64,baha}. Subsequently, 
by including the neutron-proton degree of freedom it was extended to quartets of nuclei, 
in which an even-even, an odd-proton, an odd-neutron and an odd-odd nucleus form a 
supermultiplet \cite{quartet}. Supersymmetry imposes strong constraints on both the 
collective (bosonic) and the single-particle (fermionic) degrees of freedom. 
Nevertheless, various nuclei have been identified as examples. The odd-odd nucleus 
$^{196}$Au, together with the odd-neutron nucleus $^{195}$Pt, the odd-proton nucleus 
$^{195}$Au and the even-even nucleus $^{194}$Pt, have been verified experimentally 
using state-of-the-art techniques \cite{metz,groeger,wirth} to closely fulfill 
the rules that define a supersymmetric quartet \cite{quartet,jolie}. The interpretation 
of these four nuclei as members of a supersymmetric quartet made it possible to predict 
\cite{quartet} the properties of $^{196}$Au almost 15 years before they were measured 
experimentally \cite{metz,groeger}. 

It is the purpose of this paper to present evidence for the presence of a new quartet 
of supersymmetric nuclei in the mass $A \sim 190$ region, consisting of the $^{192,193}$Os 
and $^{193,194}$Ir nuclei. The evidence is based both on energies and transfer strengths. 

\section{Supersymmetric quartet of nuclei}

The $A \sim 190$ region of the nuclear mass table is a particularly complex 
one, displaying transitional behavior such as prolate-oblate deformed shapes, 
$\gamma$-unstability, triaxial deformation and/or coexistence of different 
configurations which present a daunting challenge to nuclear structure models. 
Despite this complexity, the $A \sim 190$ mass region has been a rich ore   
of empirical evidence for the existence of dynamical symmetries and supersymmetries 
in nuclei both for even-even, odd-proton, odd-neutron and odd-odd nuclei, as well as 
for supersymmetric pairs \cite{FI,u64,baha} and quartets of nuclei 
\cite{quartet,metz,groeger,wirth,barea}. In addition to providing a unified 
description of collective nuclei, the dynamical (super)symmetries of the IBM and its 
extensions provide a powerful tool to unravel and classify the spectra of complex 
nuclei by means of a set of closed expressions for energies, electromagnetic 
transition rates and spectroscopic factors for one- and two-nucleon transfer reactions, 
which can be used to analyze, classify and interpret the experimental data. 
As an example, we mention the interpretation of the $^{194,195}$Pt and $^{195,196}$Au 
nuclei as members of a supersymmetric quartet with $U(6/12)_{\nu} \otimes U(6/4)_{\pi}$ 
supersymmetry \cite{quartet,metz}, in which the odd proton is allowed to occupy the 
$2d_{3/2}$ orbit of the 50-82 shell, and the odd neutron the $3p_{1/2}$, $3p_{3/2}$ 
and $2f_{5/2}$ orbits of the 82-126 shell. This supermultiplet is characterized by 
${\cal N}_{\pi}=2$ and ${\cal N}_{\nu}=5$. In this scheme, the excitation spectra of 
the four nuclei that constitute a quartet are described simultaneously by a single 
energy formula 
\ba
E &=& A \left[ N_1(N_1+5)+N_2(N_2+3)+N_3(N_3+1) \right] 
\nonumber\\
&& + B  \left[ \Sigma_1(\Sigma_1+4)+\Sigma_2(\Sigma_2+2)+\Sigma_3^2 \right] 
\nonumber\\
&& + B' \left[ \sigma_1(\sigma_1+4)+\sigma_2(\sigma_2+2)+\sigma_3^2 \right] 
\nonumber\\
&& + C \left[ \tau_1(\tau_1+3)+\tau_2(\tau_2+1) \right] 
\nonumber\\
&& + D \, L(L+1) + E \, J(J+1) ~,
\label{npsusy} 
\ea  
using the same values of the coefficients $A$, $B$, $B'$, $C$, $D$ and $E$ for all 
four nuclei. The first three terms in Eq.~(\ref{npsusy}) correspond to vibrational 
excitations and the final three to rotations.  

Recently, the structure of the odd-odd nucleus $^{194}$Ir was investigated by a series 
of transfer and neutron capture reactions \cite{balodis}. 
The odd-odd nucleus $^{194}$Ir differs from $^{196}$Au by two protons, the 
number of neutrons being the same. The latter is crucial, since the dominant 
interaction between the odd neutron and the core nucleus is of quadrupole type, 
which arises from a more general interaction in the IBFM for very special values 
of the occupation probabilities of the $3p_{1/2}$, $3p_{3/2}$ and $2f_{5/2}$ 
orbits, {\em i.e.} to the location of the Fermi surface for the neutron orbits 
\cite{bijker}. This situation is satisfied to a good approximation by the 
$^{195}$Pt and $^{196}$Au nuclei which both have the 117 neutrons. The same is 
expected to hold for the isotones $^{193}$Os and $^{194}$Ir. In particular, the new data 
from the polarized $(\vec{d},\alpha)$ transfer reaction provided crucial new information 
about and insight into the structure of the spectrum of $^{194}$Ir which led 
to significant changes in the assignment of levels as compared to previous work 
\cite{joliegarrett}. Theoretically, the levels of this nucleus were interpreted 
successfully in terms of a dynamical symmetry in odd-odd nuclei \cite{balodis}. 

\begin{figure}
\centering
\includegraphics[width=0.9\columnwidth]{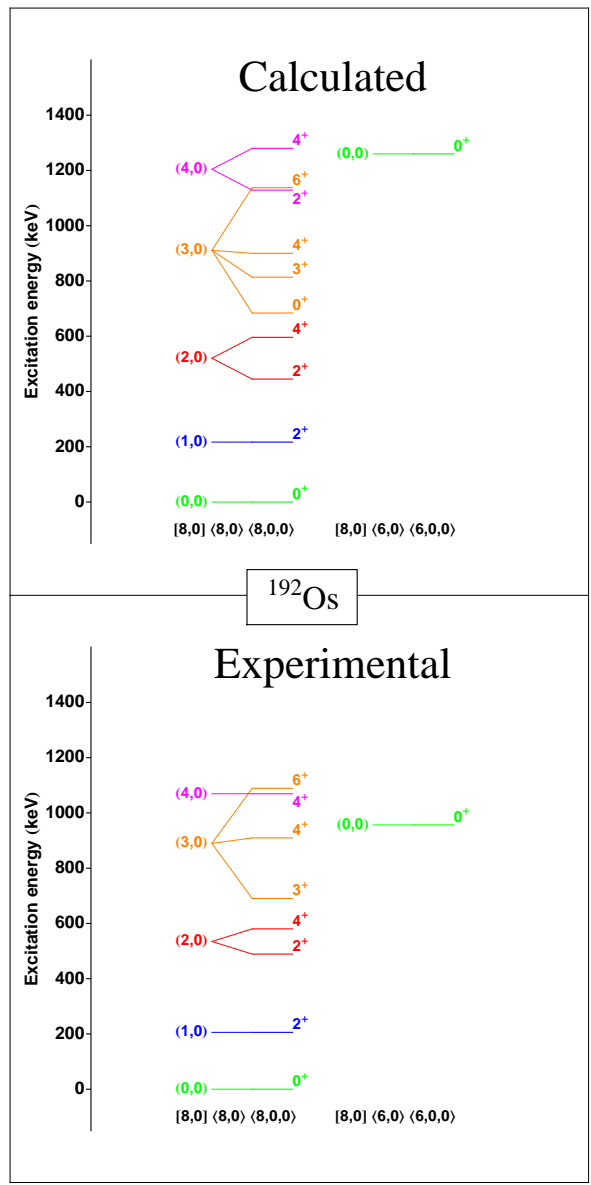} 
\caption[]{(Color online) Comparison between the theoretical and experimental spectrum of 
the even-even nucleus $^{192}$Os. The theoretical spectrum is calculated for the 
$U_{\nu}(6/12)\otimes U_{\pi}(6/4)$ supersymmetry scheme with Eq.~(\ref{npsusy}). 
The parameter values are given in Table~\ref{parameters}.}
\label{osmium92} 
\end{figure}

\begin{figure}
\centering
\includegraphics[width=0.9\columnwidth]{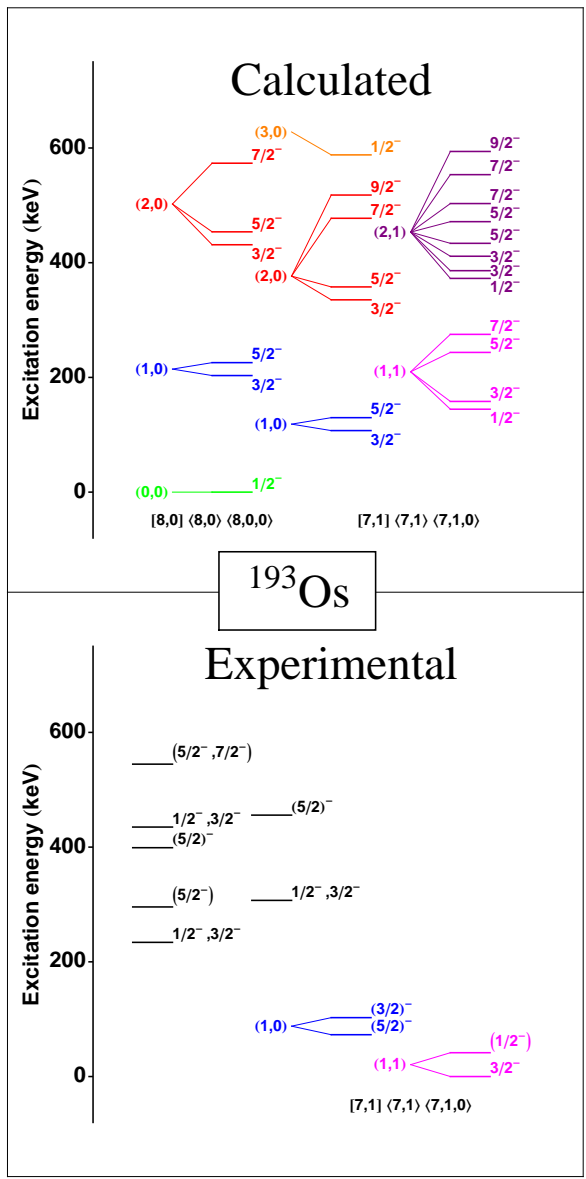} 
\caption[]{(Color online) As Fig.~\ref{osmium92}, but for the odd-neutron nucleus $^{193}$Os.}
\label{osmium93} 
\end{figure}

The successful description of the odd-odd nucleus $^{194}$Ir opens the possibility 
of identifying a second quartet of nuclei in the $A \sim 190$ mass region with 
$U(6/12)_{\nu} \otimes U(6/4)_{\pi}$ supersymmetry. The new quartet consists of the 
nuclei $^{192,193}$Os and $^{193,194}$Ir and is characterized by ${\cal N}_{\pi}=3$ 
and ${\cal N}_{\nu}=5$. The energy spectra of the quartet of nuclei are described 
simultaneously by the energy formula of Eq.~\ref{npsusy} with a single set of parameters. 
The pair of nuclei $^{192}$Os-$^{193}$Ir has been studied as an example of $U(6/4)$ 
supersymmetry by Bars, Balantekin and Iachello \cite{u64} who found that the rotational 
levels in both nuclei can be described to a good approximation by $C=40$ keV and $D+E=10$ keV.  
The $J^P=0^+$ state at 1206 keV in the even-even nucleus $^{192}$Os is interpreted 
as a vibrational excitation, leading to $B+B'=-33.5$ keV, whereas the $J^P=\frac{3}{2}^+$ 
state at 460 keV in the odd-proton nucleus $^{193}$Ir has been identified as the 
bandhead of a vibrational excitation, which gives $B'=-25.5$ keV. More recently, 
the odd-odd nucleus $^{194}$Ir was studied both experimentally and theoretically 
in Ref.~\cite{balodis}. The rotational levels were described by $C=35.1$ keV, 
$D=6.3$ keV and $E=4.5$ keV, in good agreement with the values determined previously 
for $^{192}$Os-$^{193}$Ir. The available experimental information on $^{194}$Ir 
allowed to determine two of the three vibrational terms, $A+B=35$ keV and $B'=-33.6$ keV. 
Finally, in absence of detailed experimental information on the odd-neutron nucleus 
$^{193}$Os, this nucleus was not taken into account in the fit. A simultaneous fit to 
the energy levels in $^{192}$Os, $^{193}$Ir and $^{194}$Ir with the energy formula of 
Eq.~(\ref{npsusy}) gives $A = 41.0$, $B = -6.0$, $B' = -29.0$, $C = 38.0$, $D = 6.3$ and 
$E = 4.5$ (all in keV). 

\begin{table}
\centering
\caption[]{Values of the parameters in keV}
\label{parameters}
\begin{tabular}{crrrrrrc}
\hline
& $A \;$ & $B \;$ & $B' \;$ & $C \;$ & $D \;$ & $E \;$ & Ref. \\
\hline
Os-Ir & 41.0 & --6.0 & --29.0 & 38.0 &   6.3 &  4.5 & Present \\
Os-Ir & 63.0 & --9.3 & --24.2 & 36.1 & --5.1 & 15.9 & \cite{joliegarrett} \\
Pt-Au & 52.5 &   8.7 & --53.9 & 48.8 &   8.8 &  4.5 & \cite{groeger} \\
\hline 
\end{tabular}
\end{table}

The main difference with the parametrization of \cite{joliegarrett} is due to the new 
experimental information on the odd-odd nucleus $^{194}$Ir, which has led to an interchange 
in the assignments of the ground band and the first excited bands in $^{194}$Ir \cite{balodis}. 
The fitted values of $B$, $B'$, $C$ and $D+E$ are essentially the same, since in both studies 
they are determined from the energy spectra of the pair of nuclei $^{192}$Os-$^{193}$Ir.   
The difference in the values of $A$ and $D$ (or $E$) arises from the fact that in 
\cite{joliegarrett} they were extracted from the (scarce) experimental information on the 
odd-neutron nucleus $^{193}$Os, whereas in the present study we used the new detailed 
experimental data on the odd-odd nucleus $^{194}$Ir to determine their values. 
In addition, the present parameter set is closer to the parameter values determined 
for the quartet $^{194,195}$Pt-$^{195,196}$Au \cite{groeger} than \cite{joliegarrett} 
(see Table~\ref{parameters}), indicating systematics in this zone of the nuclear chart. 

\begin{figure}
\centering
\includegraphics[width=0.9\columnwidth]{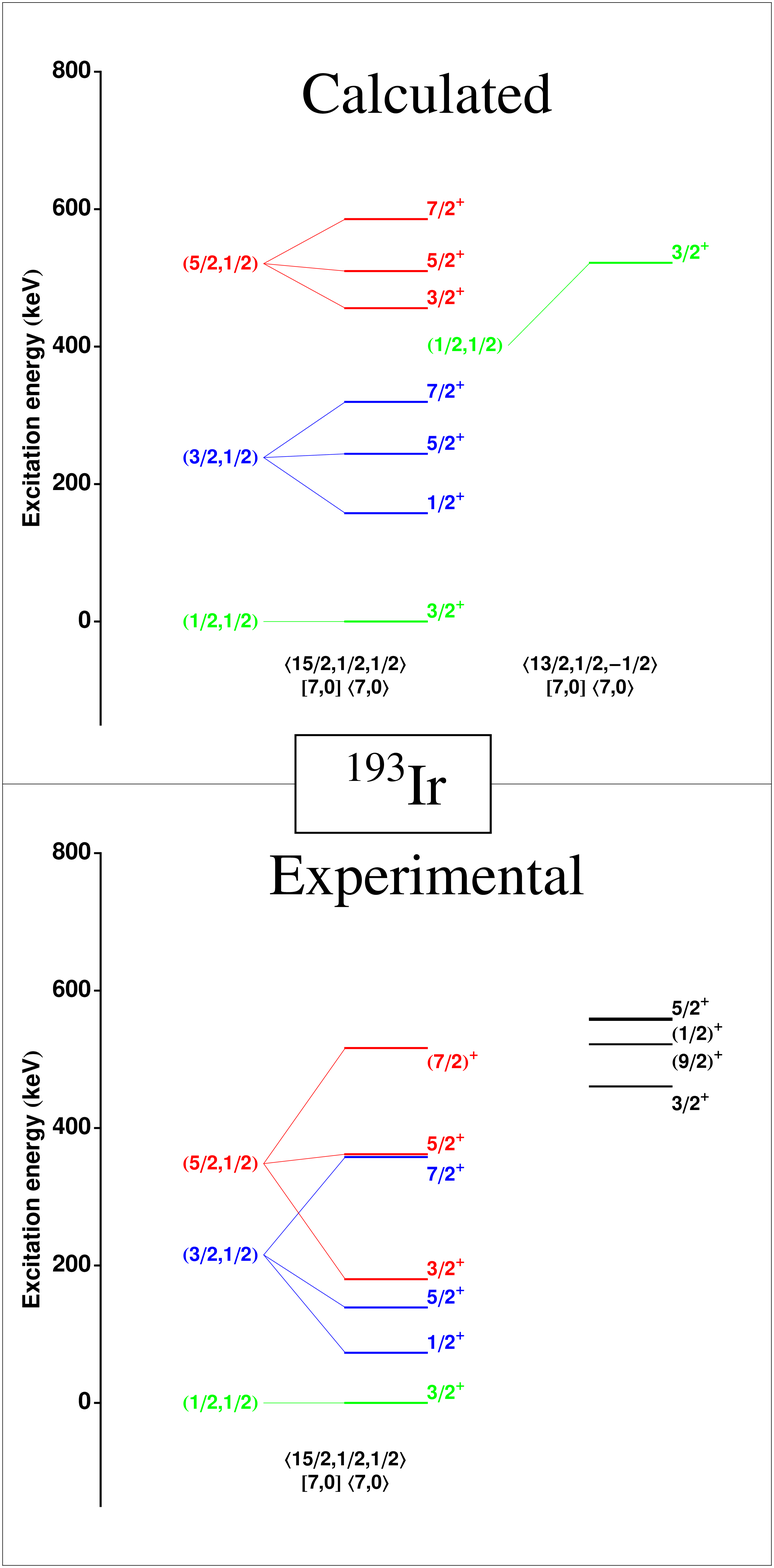} 
\caption[]{(Color online) As Fig.~\ref{osmium92}, but for the odd-proton nucleus $^{193}$Ir.}
\label{ir193} 
\end{figure}

\begin{figure}
\centering
\includegraphics[width=0.9\columnwidth]{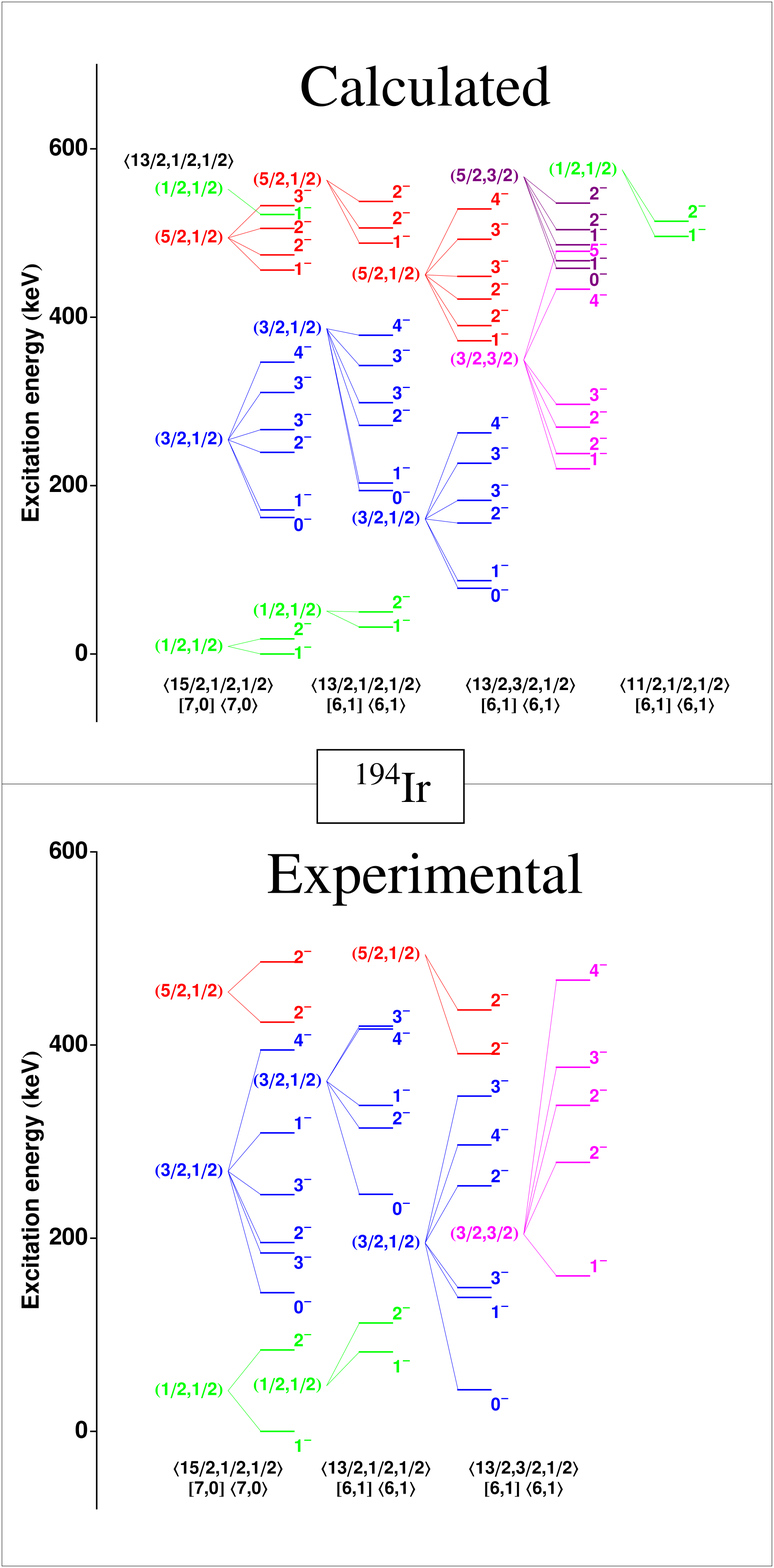} 
\caption[]{(Color online) As Fig.~\ref{osmium92}, but for the odd-odd nucleus $^{194}$Ir.}
\label{ir194} 
\end{figure}

In Figs.~\ref{osmium92}-\ref{ir194}, we show a comparison between the experimental 
and theoretical spectra for the quartet of nuclei $^{192,193}$Os-$^{193,194}$Ir    
in the $U(6/12)_{\nu} \otimes U(6/4)_{\pi}$ supersymmetry scheme. Given the complex 
nature of the spectrum of heavy nuclei in the mass $A \sim 190$ region, and in 
particular that of the odd-odd nuclei, the agreement is remarkable. There is an 
almost one-to-one correlation between the experimental and theoretical level 
schemes. Whereas the even-even nucleus $^{192}$Os, the odd-proton nucleus $^{193}$Ir 
and the odd-odd nucleus $^{194}$Ir are well-known experimentally, the data on the 
odd-neutron nucleus $^{193}$Os are rather scarce. In Fig.~\ref{osmium93} we show the 
predicted spectrum for $^{193}$Os as obtained from Eq.~(\ref{npsusy}) using the parameter 
set determined from a fit to the nuclei $^{192}$Os and $^{193,194}$Ir. 

The ground state of $^{193}$Os has spin and parity $J^P=\frac{3}{2}^{-}$, which implies 
that the second band with labels $[7,1]$, $\langle 7,1,0 \rangle$ is the ground state band, 
rather than $[8,0]$, $\langle 8,0,0 \rangle$. The assignment of levels in Fig.~\ref{osmium93} 
is based in part on preliminary data from the $^{192}$Os$(\vec{d},p){}^{193}$Os  
one-neutron transfer reaction \cite{eisermann} which shows that the $j=\frac{1}{2}$ 
strength goes to the state at 234 keV, whereas the $j=\frac{3}{2}$ and $j=\frac{5}{2}$ 
transfers are predominantly to the states at 103 and 73 keV, respectively. 
The $J^P=\frac{3}{2}^{-}$ ground state is populated very weakly, whereas the first excited 
state at 41 keV is not seen at all in this reaction. 

Theoretically, these transfer reactions are described by the fermionic generators of 
the superalgebra which change a boson into a fermion. In a study of the stripping reaction 
$^{194}$Pt $\rightarrow$ $^{195}$Pt in which the initial and final nucleus have the same 
number of neutrons as for $^{192}$Os $\rightarrow$ $^{193}$Os, it was found that the 
intensities for $j=\frac{1}{2}$ transfers are described by the operator \cite{bi} 
\begin{eqnarray}
P_{\nu}^{(\frac{1}{2}) \, \dagger} &=& \frac{\alpha_{\frac{1}{2}}}{\sqrt{6}} 
\left[ \left( \tilde{s}_{\nu} a^{\dagger}_{\nu,\frac{1}{2}} \right)^{(\frac{1}{2})} 
- \sqrt{2} \left( \tilde{d}_{\nu} a^{\dagger}_{\nu,\frac{3}{2}} \right)^{(\frac{1}{2})} \right.
\nonumber\\ 
&& \hspace{2cm} \left.
+ \sqrt{3} \left( \tilde{d}_{\nu} a^{\dagger}_{\nu,\frac{5}{2}} \right)^{(\frac{1}{2})} \right] ~. 
\label{p0}
\end{eqnarray}
According to the seleccion rules, only the $J^P=\frac{1}{2}^{-}$ state with 
$(\tau_1,\tau_2)=(0,0)$, $L=0$ of the symmetric band with $[8,0]$, $(8,0,0)$ are populated. 
This suggests to identify the energy level at 234 keV,with this state. 

The one-neutron $j=\frac{3}{2}$, $\frac{5}{2}$ transfer are described by the operators 
\begin{eqnarray}
P_{\nu}^{(j) \, \dagger} &=& \frac{\alpha_j}{\sqrt{2}} 
\left[ ( \tilde{s}_{\nu} a^{\dagger}_{\nu,j} )^{(j)} 
- ( \tilde{d}_{\nu} a^{\dagger}_{\nu,\frac{1}{2}} )^{(j)} \right] ~.
\label{p2}
\end{eqnarray}
This operator can excite the $J^P=\frac{3}{2}^{-}$, $\frac{5}{2}^{-}$ 
doublets with $(\tau_1,\tau_2)=(1,0)$, $L=2$ belonging to the bands with $[8,0]$, $(8,0,0)$ and 
$[7,1]$, $(7,1,0)$  (see Figures~\ref{osmium92}-\ref{osmium93}). Since ratios of intensities do not depend 
on the value of the coefficient $\alpha_j$ and provide a direct test of the wave functions, 
we consider the ratio $R$ for the excitation of the doublet of the band with $[{\cal N}-1,1]$, 
$({\cal N}-1,1,0)$ relative to that of the symmetric band $[{\cal N},0]$, $({\cal N},0,0)$ 
\cite{bi} 
\begin{eqnarray}
R({\rm ee \rightarrow on}) = \frac{({\cal N}-1)({\cal N}+1)({\cal N}+4)}{2({\cal N}+2)} ~,
\label{ratio}
\end{eqnarray}
which gives $R=37.8$ for the stripping reaction $^{192}$Os $\rightarrow$ $^{193}$Os 
(with ${\cal N}={\cal N}_{\pi}+{\cal N}_{\pi}=8$), 
{\it i.e.} most of the strength goes to the doublet with $[7,1]$, $(7,1,0)$. For this reason, 
we assign the states at 103 and 73 keV as the $(\tau_1,\tau_2)=(1,0)$, $L=2$ doublet of the  
$[7,1]$, $(7,1,0)$ band. 

Finally, the ground state and the first excited state of $^{193}$Os are assigned as members 
of a $J^P=\frac{3}{2}^{-}$, $\frac{1}{2}^{-}$ doublet with $(\tau_1,\tau_2)=(1,1)$, $L=1$ 
of the $[7,1]$, $(7,1,0)$ band, since neither of these states can be excited by the transfer 
operators of Eqs.~(\ref{p0}) and (\ref{p2}).  

In order to establish the assignments of the energy levels of $^{193}$Os on a firmer basis 
and to test the predictions for their doublet structure of either nuclear supersymmetry and/or 
the particle-triaxial rotor model \cite{petkov}, it is of great importance that the nucleus 
$^{193}$Os be studied in more detail experimentally. 

\section{Summary and conclusions}

In conclusion, symmetries and supersymmetries play an important role to analyze, 
classify, interpret and understand the spectra of complex many-body quantum systems. 
In this manuscript, we presented evidence for the existence of a second quartet 
of nuclei in the mass $A \sim 190$ region with $U_{\nu}(6/12)\otimes U_{\pi}(6/4)$ 
supersymmetry, consisting in the $^{192,193}$Os and $^{193,194}$Ir nuclei. 
An analysis of the energy spectra of the four nuclei that make up the quartet 
shows that the parameter set obtained in 1981 for the pair $^{192}$Os-$^{193}$Ir
\cite{u64} is very close to that of $^{194}$Ir \cite{balodis}, which indicates  
that the nuclei $^{192,193}$Os and $^{193,194}$Ir may be interpreted in terms of  
a quartet of nuclei with $U(6/12)_{\nu} \otimes U(6/4)_{\pi}$ supersymmetry. 
The supersymmetry in this new quartet of nuclei is satisfied with an accuracy comparable, 
if not better, to that found in the $^{194,195}$Pt and $^{195,196}$Au nuclei, which 
was theoretically predicted almost 25 years ago and confirmed in 1999. 

Nuclear supersymmetry establishes precise links among the spectroscopic properties of 
different nuclei, a fact that has been used in this Rapid Communication to predict 
the energy spectrum of the odd-neutron nucleus $^{193}$Os from the known properties 
of the remaining three nuclei that make up the quartet. Similar relations hold for 
other observables, such as electromagnetic and transfer strengths. Since the wave 
functions of the members of a supermultiplet are connected by symmetry, there exists 
a high degree of correlation between different one- and two-nucleon transfer reactions 
not only between nuclei belonging to the same multiplet \cite{JPA}, but also for nuclei 
from different quartets. As an example of the latter, we are currently considering a 
set of two-proton transfer experiments between different pairs of nuclei in the two 
quartets of the Os-Ir and Pt-Au nuclei \cite{twoproton}. 

\begin{acknowledgments}
This work was supported in part by PAPIIT-UNAM (grant IN113808), 
and in part by the Deutsche Forschungsgemeinschaft (grants JO391/2-3 and GR894/2-3). 
\end{acknowledgments}

\end{document}